\documentstyle[12pt]{article}
\begin{document}
\begin{titlepage}

\begin{center}
{\LARGE  Strings with interacting ends$^{\star}$}

\vskip 1cm
{\bf P.W\c{e}grzyn$^{^{\dag}}$ } \\
\vskip 1cm

\end{center}

\begin{abstract}

At the classical level we study open bosonic strings.
A generic description of
string self--interactions localized
at string ends is given. Self--interactions are characterized
by two dimensionless coupling constants. The model is
rewritten using complex Liouville fields. Using these Lorentz and
reparametrization invariant variables, equations of motion
get greatly simplified and reduce to some boundary problem
for Liouville equation.

\end{abstract}

\vskip 2cm
\begin{tabular}{l}
TPJU--23/94 \\
December 1994
\end{tabular}

\vspace{\fill}

\noindent
\underline{\hspace*{10cm}}

\noindent
$^{\star}$Work supported in part by grant KBN 2 P302 049 05.

\noindent
$ ^{\dag}$Permanent address: Department of Field Theory,
 Institute of Physics,\\
Jagellonian University, Reymonta 4, 30--059 Cracow, Poland.\\
{\tt e-mail: wegrzyn@ztc386a.if.uj.edu.pl}

\noindent
\end{titlepage}

The classical dynamics of relativistic strings follows entirely from
the minimal action principle. The action integral is some
reparametrization invariant functional of time--like two--dimensional
surfaces. For open strings, equations of motion derived from
the action consist of two groups, bulk and boundary equations.
Bulk equations of motion are Euler--Lagrange variational equations.
In the simplest case of non-interacting Nambu--Goto strings they
are Laplace--Beltrami equations and their solutions represent
minimal surfaces, i.e. (time--like) surfaces of zero mean
external curvature. For open Nambu--Goto strings,
we know that Laplace--Beltrami equations should be
supplemented by (von Neumann type) boundary conditions
for world sheet coordinates. In general, if we refer to any would-be
string model, boundary conditions to be
satisfied at open string ends have a form of  dynamical equations.
They can describe some kind of string self--interactions.
In the paper, we want to follow this point through. We believe
that the problem is essential while considering string models
of hadrons. The description of the hadronic string should have
regard to the presence of quarks at the ends, so that
it is instructive to examine possible interactions
between the string and its ends.

Let us begin with the introduction of the generic form of the
action functional for open strings.
\begin{equation}
S = \int_{\tau_{1}}^{\tau_{2}} \ d\tau
\int_{\sigma_{1}(\tau)}^{\sigma_{2}(\tau)} \ d\sigma \
\sqrt{-g} {\cal L} +
  \int_{\tau_{1}}^{\tau_{2}} \ d\tau \ L^{(1)} +
  \int_{\tau_{1}}^{\tau_{2}} \ d\tau \ L^{(2)} \ ,
\label{action}
\end{equation}
where ${\cal L}$ is a scalar function (with respect to both
Poincare and reparametrization transformations) of string
coordinates. It can be always presented as a functional
of induced metric
$g_{ab} = X^{\mu}_{,a} X_{\mu,b}$ and defined with its help
covariant derivatives of world sheet coordinates
$X_{\mu}(\sigma^{a})$ ($\sigma^{a} = (\tau,\sigma)$):
\begin{equation}
{\cal L} = {\cal L} (g^{ab}; \nabla_{a} X_{\mu};
\nabla_{a}\nabla_{b} X_{\mu}; ...) \ ,
\end{equation}

$$ \nabla_{a} X_{\mu} \equiv \frac{\partial X_{\mu}}{\partial
\sigma^{a}} \equiv X_{\mu,a} \ , \ \
\nabla_{a} \nabla_{b} X_{\mu} \equiv X_{\mu,ab}
- \Gamma^{c}_{ab} X_{\mu,c} \ .$$
The dots stand for higher covariant derivatives, additional
fields placed on the world sheet or external fields
coupled with strings. Lagrangians $L^{(i)}$ are functionals
of string ends trajectories and their total time derivatives:
\begin{equation}
L^{(i)} = L^{(i)} (X_{\mu}; d_{t} X_{\mu}; ...)
\big|_{\sigma = \sigma_{i}(\tau)} \ ,
\end{equation}

$$ d_{t} = \nabla_{0} + \dot{\sigma_{i}} \nabla_{1} \ .$$
Similarly, the dots stand here for higher total time derivatives
and couplings with external fields. Poincare--invariant
functionals $L^{(i)}$ should be also scalar densities with respect
to the change of parametrization of string ends trajectories
(such transformations can be a part of any total world sheet
reparametrization).

The bulk classical equations of motion derived  from
the action integral (\ref{action}) follow entirely from
the first action term. They can be written down
in a manifestly covariant form \cite{pw}:
\begin{equation}
\sqrt{-g} \nabla_{a} \Pi^{a}_{\mu} = 0 \ ,
\end{equation}
where $\Pi^{a}_{\mu}$ is given by the formula
\begin{eqnarray}
\Pi^{a}_{\mu} = - {\cal L} \nabla^{a} X_{\mu}
- \frac{\partial {\cal L}}{\partial X^{\mu}_{,a}} +
2 \frac{\partial {\cal L}}{\partial g^{bc}} g^{ab} \nabla^{c} X_{\mu}
\nonumber \\
+ \nabla_{b} \left[ \frac{\partial {\cal L}}{\partial
(\nabla_{a} \nabla_{b} X^{\mu})} \right] \ .
\end{eqnarray}
{}From now, we restrict our discussion to string models
defined by Lagrangians that depend on no higher than
second order derivatives.
The above formula is much more simpler to evaluate Euler--Lagrange
equations than the standard one which includes variational derivatives
taken with respect to non--covariant world sheet derivatives.
As usual, performing variational derivatives we regard $g^{01}$
and $g^{10}$, $\nabla_{0}\nabla_{1}X_{\mu}$ and
$\nabla_{1}\nabla_{0}X_{\mu}$ as independent variables. Then, all
variational derivatives are tensor objects.

For open strings, the variational problem results also
in 'the edge conditions', being in fact dynamical
equations of motion to be held at world sheet boundaries
(trajectories of string end points).
The boundary equations of motion are collected below
\begin{eqnarray}
\sqrt{-g} \Pi^{1}_{\mu} +
\partial_{0} \left[ \sqrt{-g}
\frac{\partial {\cal L}}{\partial (\nabla_{0} \nabla_{1} X^{\mu})}
\right]
- \partial_{1} \left[ \sqrt{-g}
\frac{\partial {\cal L}}{\partial (\nabla_{0} \nabla_{0} X^{\mu})}
\right]  \dot{\sigma}_{i}^{2} \nonumber \\
- \left\{ \sqrt{-g} \Pi^{0}_{\mu} +
\partial_{0} \left[ \sqrt{-g}
\frac{\partial {\cal L}}{\partial (\nabla_{0} \nabla_{0} X^{\mu})}
\right]
- \partial_{1} \left[ \sqrt{-g}
\frac{\partial {\cal L}}{\partial (\nabla_{0} \nabla_{1} X^{\mu})}
\right] \right\} \dot{\sigma}_{i} \nonumber \\
- \sqrt{-g}
\frac{\partial {\cal L}}{\partial (\nabla_{0} \nabla_{0} X^{\mu})}
\ddot{\sigma}_{i} +
(-1)^{i} d_{t} \left( \frac{\partial  L}{\partial d_{t}
X^{\mu}} \right)
- (-1)^{i} d^{2}_{t} \left( \frac{\partial  L}{\partial
d^{2}_{t} X^{\mu}} \right) = 0 \ ,
\label{b1}
\end{eqnarray}

\begin{eqnarray}
\sqrt{-g}
\frac{\partial {\cal L}}{\partial (\nabla_{1} \nabla_{1} X^{\mu})}
- \left[ \sqrt{-g}
\frac{\partial {\cal L}}{\partial (\nabla_{0} \nabla_{1} X^{\mu})}
+ \sqrt{-g}
\frac{\partial {\cal L}}{\partial (\nabla_{1} \nabla_{0} X^{\mu})}
\right] \dot{\sigma}_{i} \nonumber \\
+ \sqrt{-g}
\frac{\partial {\cal L}}{\partial (\nabla_{0} \nabla_{0} X^{\mu})}
\dot{\sigma}_{i}^{2} = 0 \ , \ \ \ \
{\rm for} \ \sigma=\sigma_{i}(\tau) \ .
\label{b2}
\end{eqnarray}

In the ordinary Nambu--Goto model ${\cal L}$ is a constant
and 'point-like' terms in (\ref{action}) are absent.
The Nambu--Goto string action can be considered as the
action of a non-interacting string.
Any kind of string self--interactions must involve
higher derivatives of world sheet coordinates in the
action. In this paper, we consider only such interactions
that leave bulk equations of motion unchanged
(i.e. the same as in the Nambu--Goto case):
\begin{equation}
\label{ng}
g^{ab} \nabla_{a} \nabla_{b} X_{\mu} = 0 \ .
\end{equation}
Thus, possible string self--interactions are localized
at string ends.
A general form of the action that results in the Nambu--Goto
bulk string equations has been derived in \cite{pw}:
\begin{equation}
\label{lagr}
{\cal L} = - \gamma - \frac{\alpha}{2} R
- \beta  N   \ ,
\end{equation}
where $\gamma$ is string tension, $\alpha$ and $\beta$ are
dimensionless parameters.
 The first constant term in (\ref{lagr}) corresponds to
the Nambu--Goto action, while other terms are respectively
the integrands of Gauss--Bonnet and Chern invariants for
two--dimensional surfaces. Scalar functions $R$ and $N$
have the form
\begin{equation}
R = (g^{ab} g^{cd} - g^{ad} g^{bc})
(\nabla_{a} \nabla_{b} X_{\mu})
\nabla_{c} \nabla_{d} X^{\mu} \ ,
\end{equation}
\begin{equation}
\label{ns}
N = - \frac{1}{2 \sqrt{-g}} g^{ac} \in^{bd} \tilde{t}^{\mu \nu}
(\nabla_{a} \nabla_{b} X_{\mu})
\nabla_{c} \nabla_{d} X_{\nu} \ ,
\end{equation}
where $\tilde{t}^{\mu \nu} = \frac{1}{\sqrt{-g}} \in^{\mu \nu
\rho \sigma}  X_{\rho,0} X_{\sigma,1}$.

For point--like terms in the action (\ref{action}), we restrict
ourselves to the simplest choice and take invariant lengths
of trajectories of string ends [2--4], namely
\begin{equation}
L^{(i)} = - m_{i} \sqrt{(d_{t} X)^{2}}
\Big|_{\sigma=\sigma_{i}(\tau)} \ .
\label{lagr2}
\end{equation}
Physically, we can say that point--like masses $m_{1}$ and $m_{2}$
have been attached to string ends.

A great technical advantage of working with Nambu--Goto strings is
that bulk equations of motion (\ref{ng}) can be linearized
by a suit choice of gauge. Practically, any kind of string
self--interactions that changes bulk equations (\ref{ng})
complicates them in such a drastic way that makes the model
mathematically hardly tractable (we have to deal with a non--linear
system of partial differential equations of fourth order in time
and space derivatives). The purpose of this paper is to establish
a suitable formalism to examine string self--interactions that
result merely in boundary equations of motion. It is astonishing
that if we try to construct invariant action terms for this type
of self--interactions, then we find only two possibilities (displayed
in (\ref{lagr})). To prove this fact (see Ref.\cite{pw}) we
need only consider requirements of Poincare and reparametrization
invariances. It is important to stress that in this proof
{\em no} assumptions are made on analytical form of invariant
terms. This makes our statement really strong. For example,
let us compare it with another well--known statement that
scalar functions $R$, $N$ and a square of Laplace--Beltrami
operator (Polyakov rigidity term) are the only invariants we
can construct for two--dimensional surfaces immersed in
four--dimensional flat spacetime. This statement is true as
long as we restrict ourselves to action terms for surfaces
such that their integrands are polynomials in external curvature
tensor coefficients and respective coupling constants have
 "renormalizable" dimensions. If we weaken any of these
assumptions, we find immediately infinitely many new candidates
for action terms.

Let us start our treatment of the model defined by the action
(\ref{action}), with Lagrangians specified
by (\ref{lagr}) and (\ref{lagr2}). As it has been explained,
string Lagrangian (\ref{lagr}) is the most general form
allowed by a restriction that the variational problem results
in Nambu--Goto bulk equations of motion (\ref{ng}), and
point--like Lagrangian (\ref{lagr2}) is just the simplest
choice. However, it is the only possible choice unless we let
higher than third order derivatives appear in boundary
equations. Boundary equations of motion (\ref{b1}--\ref{b2})
arise from string self--interaction terms in (\ref{lagr})
and point--like Lagrangians (\ref{lagr2}). These equations
are pretty complicated non--linear differential equations
with third derivatives in time and space parameters if they are
expressed in terms of world sheet coordinates $X_{\mu}$.
They are to be held at both string ends $\sigma = \sigma_{i}$
at any time.

We will make use of the following convenient way of
parametrizating string world sheets \cite{bn}
\begin{eqnarray}
(X_{\mu,0} \pm X_{\mu,1})^{2} = 0 \ , \nonumber \\
(X_{,00} \pm X_{,01})^{2}  = - q^{2} \ ,
\label{gauge}
\end{eqnarray}
where $q$ is some arbitrary constant of mass dimension.
The first set of parametrization conditions (orthonormal gauge)
makes bulk equations of motion (\ref{ng}) linear
and their general solution can
be represented as the combination of left-- and right--moving parts,
\begin{equation}
\label{sol}
X_{\mu}(\tau,\sigma) = X_{L\mu}(\tau+\sigma) + X_{R\mu}(\tau-\sigma)
\ .
\end{equation}
As the orthonormal gauge still allows for
conformal changes of parametrization,
so that the latter pair of parametrization conditions in (\ref{gauge})
is chosen to kill completely this residual
gauge freedom.  Note that this supplementary gauge is possible
only for world sheets obeying bulk equations of motion.

Up to transformations of Poincare group, minimal surfaces $X_\mu$
(i.e. surfaces satisfying bulk equations (\ref{ng})
parametrized according to (\ref{gauge})) correspond to solutions
$\Phi$ of the complex Liouville equation \cite{pw}:
\begin{equation}
\label{liouv}
\ddot{\Phi} - \Phi'' = 2q^{2}e^{\Phi}.
\end{equation}
The real part of Liouville field ${\rm Re} \Phi$ is the only
independent part of the induced metric in the orthonormal gauge:
\begin{equation}
\label{metl}
\sqrt{-g} = e^{-{\rm Re} \Phi} \ .
\end{equation}
While ${\rm Re} \Phi$ describes fully internal geometry of the
string world sheet, the imaginary part of Liouville field
${\rm Im} \Phi$
characterizes its extrinsic geometry, i.e. the way world sheet is
embedded in four--dimensional spacetime. In terms of geometrical
objects, we can describe this embedding by extrinsic torsion
coefficients:

$$\omega_{a} = n^{1 \mu} \partial_{a} n^{2}_{\mu} \
\ \ \ $$
where $n^{1}_{\mu}(\tau, \sigma)$ and $n^{2}_{\mu}(\tau, \sigma)$
are vectors normal to the world sheet at point $(\tau, \sigma)$.
\begin{equation}
\label{n}
n^{i \mu} X_{\mu,a} = 0 \ , \ \ n^{i \mu} n^{j}_{\mu} =
- \delta^{ij} \ .
\end{equation}
Normal vectors $n^{i}_{\mu}$ are defined modulo their
common rotation in a local plane perpendicular to the world sheet.
Such local rotation about some angle $\phi(\tau,\sigma)$
changes torsion coefficients: $\omega_{a} \rightarrow
\omega_{a} + \partial_{a} \phi$. Therefore,
the only meaningful characteristics of extrinsic geometry
is $N = - \frac{1}{\sqrt{-g}} \in^{ab} \partial_{a} \omega_{b}$.
One can convince oneself \cite{pw} that it coincides
with defined previously scalar function (\ref{ns})
and has the following
relation with the imaginary part of Liouville field:
\begin{equation}
\label{nstl}
N = \frac{q^{2}}{g} \sin{({\rm Im} \Phi)} \ .
\end{equation}

For practical purposes,  it is most convenient to express
the correspondence between world sheet coordinates and
complex Liouville field in the following way:
\begin{eqnarray}
\label{wscl}
e^{\Phi} & = & -\frac{4}{q^2}
\frac{f_{L}'(\tau+\sigma)f_{R}'(\tau-\sigma)}
     {[f_{L}(\tau+\sigma)-f_{R}(\tau-\sigma)]^2} \ , \nonumber \\
\frac{\partial}{\partial \tau} {X}^{\mu}_{L,R} & = &
\frac{q}{4|f_{L,R}'|}(1+|f_{L,R}|^2,~2Re\,f_{L,R},~2Im\,f_{L,R},
                      ~1-|f_{L,R}|^2) \ ,
\end{eqnarray}
where $f_L$ and $f_R$ are arbitrary complex functions.
The former formulae is a general solution of Liouville equation
(\ref{liouv}),
the latter is a general solution of Nambu--Goto equations (\ref{ng})
satisfying gauge conditions (\ref{gauge}).
Arbitrary functions $f_{L}$ and $f_{R}$ in both formulae
can be identified iff we put (\ref{metl}) and (\ref{nstl}).
Note that any simultaneous modular transformation of $f_{L}$ and
$f_{R}$ induces Lorentz transformation of $X_{\mu}$ while
Liouville field $\Phi$ remains invariant.

In this paper, our main purpose will be to express bulk and boundary
equations of motion for an open string, that follow
from the action (\ref{action})
with Lagrangians (\ref{lagr}) and (\ref{lagr2}), in terms
of Liouville variables.

As it was stated, instead of bulk equations of motion (\ref{ng})
there appears complex Liouville equation (\ref{liouv}).
Thus, our task is to rewrite boundary equations (\ref{b1}--\ref{b2}).
If we insert in general formulae our Lagrangians and perform
all calculations, then after using (\ref{ng}) and (\ref{gauge})
we can present boundary equations at $\sigma = \sigma_{i}(\tau)$
in the form
\begin{eqnarray}
\gamma X_{\mu,1} + \tilde{Y}_{\mu,0} +
(\gamma X_{\mu,0} + Y_{\mu,0} + \tilde{Y}_{\mu,1}) \dot{\sigma}_{i}
\nonumber \\
+ Y_{\mu,1} \dot{\sigma}^{2}_{i} + Y_{\mu} \ddot{\sigma}_{i}
+ (-1)^{i} m_{i} d_{t} \left( \frac{d_{t} X_{\mu}}{
\sqrt{ (d_{t} X)^{2} } } \right) = 0 \ ,
\label{bb1}
\end{eqnarray}
\begin{equation}
Y_{\mu} + 2 \tilde{Y}_{\mu} \dot{\sigma}_{i} +
Y_{\mu} \dot{\sigma}^{2}_{i} = 0 \ ,
\label{bb2}
\end{equation}
where we have introduced:

$$ Y_{\mu} \equiv
\frac{\alpha}{\sqrt{-g}} \nabla_{0} \nabla_{0} X_{\mu} +
\frac{\beta}{\sqrt{-g}} \tilde{t}_{\mu\nu} \nabla_{0}
\nabla_{1} X^{\nu} \ ,$$

$$ \tilde{Y}_{\mu} \equiv
\frac{\alpha}{\sqrt{-g}} \nabla_{0} \nabla_{1} X_{\mu} +
\frac{\beta}{\sqrt{-g}} \tilde{t}_{\mu\nu} \nabla_{0}
\nabla_{0} X^{\nu} \ .$$

Taking into account definitions of $Y_{\mu}$, $\tilde{Y}_{\mu}$,
$R$, $N$ and gauge conditions (\ref{gauge}) one can derive
the identities:
\begin{eqnarray}
Y^{\mu} X_{\mu,a} = \tilde{Y}^{\mu} X_{\mu,a} = 0 \ ,
\nonumber \\
Y^{\mu} X_{\mu,01} = \tilde{Y}^{\mu} X_{\mu,00} = 0 \ ,
\nonumber \\
Y^{\mu} X_{\mu,00} = - \frac{\alpha q^{2}}{2 \sqrt{-g}}
- \frac{\alpha}{4} \sqrt{-g} R - \frac{\beta}{2} \sqrt{-g} N \ ,
\nonumber \\
\tilde{Y}^{\mu} X_{\mu,01} = - \frac{\alpha q^{2}}{2 \sqrt{-g}}
+ \frac{\alpha}{4} \sqrt{-g} R + \frac{\beta}{2} \sqrt{-g} N \ .
\label{id}
\end{eqnarray}
Now, one can easily deduce from Eq.(\ref{bb2}) that
\begin{equation}
(1 + \dot{\sigma}_{i}^{2}) Y X_{,00} =
\dot{\sigma}_{i} \tilde{Y} X_{,01} = 0 \ .
\end{equation}
Combining the above equations with identities
(\ref{id}) one can obtain the following requirement:
\begin{equation}
\dot{\sigma}_{i} = 0 \  , \ \ i = 1, 2 \ .
\label{ep}
\end{equation}
It is a great technical advantage of working with gauge conditions
(\ref{gauge}) that we can put in (\ref{action}) space parameter
interval independent of time. Here, it is a consequence of
equations of motion. Of course, it was possible at the beginning,
as it is usually done, to fix say $\sigma_{1} = 0$ and
$\sigma_{2} = \pi$ by a proper reparametrization. But, it restricts
the set of allowed parametrizations and we were not sure whether
further parametrization conditions, like (\ref{gauge}), would be
acceptable. Open rigid strings provide us with an example that it may
happen for some configurations that there exists no parametrizion
which both makes $\sigma$--interval time--indendent and fulfills
orthonormality conditions $(X_{,0} \pm X_{,1})^{2} = 0$ (see
\cite{pw2}).

Now, and for the remainder of this paper, we will assume
\begin{equation}
\sigma_{1} = 0 \ , \sigma_{2} = \pi \ .
\label{ends}
\end{equation}
Note that gauge conditions (\ref{gauge})
fix world sheet parametrization
almost unique\-ly, allowing only for constant shifts of
$\tau$ and $\sigma$ parameters. Therefore, passing from
(\ref{ep}) to (\ref{ends}) we cannot fix the length
of $\sigma$-interval to be $\pi$ for all open string
configurations. It means that this constant length is an
integral of motion (related to scaling symmetry).
As parameter $q$ specified in (\ref{gauge})
was an arbitrary constant
and there is an obvious interplay between $q$ and
$\sigma_{2}-\sigma_{1}$, we choose instead that
(\ref{ends}) is fixed while $q$ will be regarded from now as
an integral of motion.

Taking into account (\ref{ends}), boundary equations of motion
(\ref{bb1}--\ref{bb2}) at string ends $\sigma = 0, \pi$
reduce to
\begin{equation}
\gamma X_{\mu,1} + \tilde{Y}_{\mu,0} +
(-1)^{i} m_{i} \frac{\partial}{\partial \tau}
\left( \frac{X_{\mu,0}}{\sqrt[4]{-g}} \right) = 0 \ ,
\end{equation}
\begin{equation}
Y_{\mu} = 0 \ .
\end{equation}
Finally, after specifying some orthonormal frame (\ref{n}) we
express covariant derivatives (normal to the world sheet)
using imaginary part of Liouville field:
\begin{equation}
\nabla_{0} \nabla_{0} X_{\mu} = q \cos{({\rm Im} \Phi/2)}
\in^{ij} t^{j} n^{i}_{\mu} \ , \ \
\nabla_{0} \nabla_{1} X_{\mu} = q \sin{({\rm Im} \Phi/2)}
t^{i} n^{i}_{\mu} \ ,
\end{equation}
where $t^{i}(\tau, \sigma)$ are some arbitrary functions
with no geometrical meaning, satisfying $t^{i} t^{i} = 1$.
Then, the boundary equations of
motion (26,27) at $\sigma = 0, \pi$ can be expressed in terms of
Liouville field:
\begin{equation}
\label{l1}
\gamma - \alpha q^{2} e^{2{\rm Re} \Phi} =
(-1)^{i} m_{i} \frac{\partial}{\partial \sigma}
\left( e^{{\rm Re} \Phi /2} \right)  \ ,
\end{equation}
\begin{equation}
\label{l2}
C \frac{\partial}{\partial \tau} {\rm Re \Phi} = 0 \ ,
\end{equation}
\begin{equation}
\label{l3}
C \cos{({\rm Im} \Phi /2)} = \beta \ ,
\end{equation}
\begin{equation}
\label{l4}
C \frac{\partial}{\partial \sigma} {\rm Im} \Phi =
(-1)^{i} 2 m_{i}
e^{-{\rm Re \Phi} /2} \cos{({\rm Im} \Phi /2)} \ ,
\end{equation}
where $C = \sqrt{\alpha^2 + \beta^2}$.
{}From (\ref{l2}) and (\ref{l3}) follow that Liouville field $\Phi$
is constant and finite (for $C \neq 0$) at boundary points.

It is seen that the analysis of classical equations of motion
has been greatly simplified due to the introduction of
Liouville fields. We end up with some boundary problem
(\ref{l1}--\ref{l4}) for Liouville equation (\ref{liouv}).
Let us make a short account of the most important points
in our construction. We have elaborated a classical model
of open strings in four--dimensional flat spacetime. The model
is defined by the action (\ref{action}) with Lagrangians
(\ref{lagr}) and (\ref{lagr2}). It is the most general string
model as long as (i) string self--interactions are assumed
to appear only in boundary equations of motion (ii) equations
of motion contain no higher than third derivatives (iii) no
additional internal fields on the world sheet or external
fields in the target space are present. The generic string action
depends on four parameters, two dimensionless coupling constants
and two masses. The variational problem for open strings
results in usual Nambu--Goto bulk equations and boundary equations,
being a set of third order differential equations. Next, we define
world sheet parametrization in a unique way (\ref{gauge}) and
introduce new independent variables (\ref{metl}) and (\ref{nstl}),
which are combined to a complex Liouville field.
Equations of motion reduce to a boundary problem for complex Liouville
equation.

Further analysis of the model of strings with interacting ends
will be made in the following papers. Some special case
has been considered in \cite{hw,ksw}.

In the last part of this paper, we discuss possible singularities
of Liouville fields. An imaginary part of $\Phi$ can be interpreted
as some angle variable, so that we are to consider singularities of
$Re \Phi$. First, let us remind the relation between
$Re \Phi$ and the determinant
of the induced ~world sheet metric (\ref{metl}).
If $Re \Phi < 0$ near some singular point, then it follows that
invariant area of world sheet piece being a neighbourhood
of this point is infinite. Under ordinary circumstances, such
solutions are not physically acceptable. Thus, singular points
can be taken into account provided that $Re \Phi > 0$
in their close vicinities.

Let us now look into the relations (\ref{wscl}). We should consider
four critical cases:

(a) $|f_{L}|$ or $|f_{R}|$ goes to infinity

(b) $|f'_{L}|$ or $|f'_{R}|$ goes to infinity

(c) $f'_{L}=0$ or $f'_{R}=0$

(c) $f_{L}$ = $f_{R}$

at some point ($\tau_{0}$,$\sigma_{0}$).

First, it is helpful to make the following observation:
$X^{\mu}_{L,0}$ or $X^{\mu}_{R,0}$ cannot vanish. If
$X^{\mu}_{L,0}=0$ at some point ($\tau_{0}$,$\sigma_{0}$),
then $X^{\mu}_{L,0}=0$ at any point ($\tau$,$\sigma$)
such that $\tau+\sigma=\tau_0+\sigma_0$. In particular,
it would vanish at some boundary point. It is impossible
as $e^{-{\rm Re} \Phi}=2X_{L,0}X_{R,0}$ and ${\rm Re} \Phi$
is finite at boundary points. Because $X^{\mu}_{L,0}$ and
$X^{\mu}_{R,0}$ are light-like vectors, it implies that their
time components cannot vanish.

Keeping in mind the above observation, we can establish that
cases (a) and (b) can be taken into account only if
ratios $f'_{L}/f_{L}^{2}$ and $f'_{R}/f_{R}^{2}$ are finite
and non-vanishing at ($\tau_{0}$,$\sigma_{0}$).
It is straightforward to show further that Liouville field
has no singularities at such points and we can easy get
rid of (a) and (b) by performing a suitable modular transformation.

In case (c), suppose that $f'_{L}(\tau_{0}+\sigma_{0})=0$.
If ${\rm Re}\Phi>0$, then $f_{L}(\tau_{0}+\sigma_{0})
 = f_{R}(\tau-\sigma)$ at any point $(\tau,\sigma)$
such that $\tau_{0}+\sigma_{0} = \tau+\sigma$.
It implies that $f_{R}$ is a constant function in some finite
interval, but we can easy convince ourselves that it is
impossible. Thus, the case (c) cannot occur, i.e. first
derivatives of $f_{L}$ and $f_{R}$ cannot take zero values.

Finally, the case (d) is the only one when we can admit
a singularity of Liouville field $\Phi$.
If $f_{L}(\tau_{0}+\sigma_{0})
 = f_{R}(\tau_0 -\sigma_0)$, then $\sqrt{-g}=0$ at
($\tau_{0}$,$\sigma_{0}$). Such string points travel with
light--like velocities.

This work was supported in part by the KBN under grant
2 P302 049 05.

\end{document}